\documentclass{aa}
\usepackage{graphicx}
\usepackage{natbib}
\usepackage{amssymb}
\usepackage{amsfonts}
\usepackage{amsbsy}
\usepackage{amsmath}

\bibpunct{(}{)}{;}{a}{}{,}

\newcommand{\oneb}{\boldsymbol{1}}
\newcommand{\ahb}{\boldsymbol{\widehat a}}

\newcommand{\ab}{\boldsymbol{a}}
\newcommand{\Ahb}{\boldsymbol{\widehat A}}

\newcommand{\Ab}{\boldsymbol{A}}
\newcommand{\Fb}{\boldsymbol{F}}
\newcommand{\eb}{\boldsymbol{e}}
\newcommand{\Cb}{\boldsymbol{C}}

\newcommand{\Hb}{\boldsymbol{H}}
\newcommand{\Ib}{\boldsymbol{I}}

\newcommand{\Nb}{\boldsymbol{N}}
\newcommand{\Nmthb}{\boldsymbol{\mathcal N}}

\newcommand{\Lambdab}{\boldsymbol{\Lambda}}
\newcommand{\Omegab}{\boldsymbol{\Omega}}
\newcommand{\Phib}{\boldsymbol{\Phi}}

\newcommand{\Pb}{\boldsymbol{P}}

\newcommand{\Smb}{\boldsymbol{\mathcal S}}
\newcommand{\Shb}{\boldsymbol{\widehat S}}
\newcommand{\Sb}{\boldsymbol{S}}

\newcommand{\Vb}{\boldsymbol{V}}
\newcommand{\Zb}{\boldsymbol{Z}}
\newcommand{\yb}{\boldsymbol{y}}
\newcommand{\Yb}{\boldsymbol{Y}}
\newcommand{\Ymb}{\boldsymbol{\mathcal Y}}

\newcommand{\wb}{\boldsymbol{w}}

\newcommand{\zerosb}{\boldsymbol{0}}
\newcommand{\onesb}{\boldsymbol{1}}

\begin{document}

   \title{A Modified ICA Approach for Signal Separation in CMB Maps}

   \author{R. Vio\inst{1}
    \and
           P. Andreani\inst{2}
          }
   \institute{Chip Computers Consulting s.r.l., Viale Don L.~Sturzo 82,
              S.Liberale di Marcon, 30020 Venice, Italy\\
              \email{robertovio@tin.it},
         \and
		  ESO, Karl Schwarzschild strasse 2, 85748 Garching, Germany\\
                  INAF-Osservatorio Astronomico di Trieste, via Tiepolo 11, 34143 Trieste, Italy\\              
		  \email{pandrean@eso.org}
             }

\date{Received .............; accepted ................}

\abstract
{}
{One of the most challenging and important problem of digital signal processing in Cosmology is
the separation of foreground contamination from cosmic microwave background (CMB). This problem becomes even more difficult in
situations, as the CMB polarization observations, where the amount of available ``a priori'' information is limited. In this case, it is necessary
to resort to the {{\it blind separation}} methods. One important member of this class is represented by the {{\it Independent Components Analysis}}
(ICA). In its original formulation, this method has various interesting characteristics, but also some limits. One of the most serious is the difficulty
to take into account any information available in advance. In particular, ICA is not able to exploit the fact that emission of CMB is the same
at all the frequencies of observations. Here, we show how to deal with this question. The connection of the proposed methodology with
the {{\it Internal Linear Composition}} (ILC) technique is also illustrated.}
{A modification of the classic ICA approach is presented and its characteristics are analyzed both analytically and by means of numerical experiments.}
{The modified version of ICA appears to provide more stable results and of better quality.}
{}
\titlerunning{ICA method}
\authorrunning{R. Vio, \& P. Andreani}
\maketitle

\section{INTRODUCTION}

The experimental progresses in the detection of cosmological emissions require a 
parallel development of data analysis techniques in order to extract the maximum physical information 
from data. In particular, different emission mechanisms are characterized by markedly distinct underlying 
physical processes. Data analysis often requires the component separation in order to study the individual characteristics.
To achieve such a goal, a link between the branch of signal processing science 
which characterizes and separates different signals and astrophysics is yet well established, and in many cases, modern signal 
processing techniques have been imported and applied in an astrophysical context. This is the case 
of the Independent Component Analysis (ICA), used for the separation of the Cosmic Microwave Background (CMB) from diffuse foregrounds 
originated by our own Galaxy \citep[see][ and references therein]{sti06}. This techniques offers several advantages. In particular, under the
only assumption of mutual statistical independence, it permits the separation of all the components contributing to an observed signal.

More formally, let the available data in the form of $N$ mean-subtracted maps $\{ \Ymb \}_{i=1}^N$, corresponding to different observing 
channels, and containing 
$M$ pixels each. If $\Yb_i = {\rm VEC}^T[\Ymb_i]$ \footnote{We recall that the operator ${\rm VEC}[\Hb]$ transforms a matrix $\Hb$ into 
a vector by stacking its columns one underneath the other. Hence, ${\rm VEC^T[\Hb]}$ provides a row array.}, these maps can be arranged in 
a $N \times M$ matrix 
\begin{equation}
\Yb = 
\left( \begin{array}{c}
\Yb_1 \\
\Yb_2 \\
\vdots \\
\Yb_N
\end{array} \right).
\end{equation} 
A common assumption in CMB observations is that each $\Ymb_i$ is given by the linear mixture (or in astrophysical term ``frequency channel'')
of $N_c$ components
$\{ \Smb_j \}_{j=1}^{N_c}$ due to different physical processes (e.g., free-free, dust re-radiation, bremsstrahlung, \ldots). In formula,
\begin{equation} \label{eq:start}
\Yb_i = \sum_{j=1}^{N_c} a_{ij} \Sb_j,
\end{equation}
with $\Sb_j = {\rm VEC}^T[\Smb_j]$ and $\{ a_{ij} \}$ constant coefficients. With this model, it is hypothesized that for the $j$th physical process 
a template $\Smb_j$ exists that is 
independent of the observing channel ``$~i~$''. Although rather strong, actually it is not unrealistic to assume that this condition is
satisfied when small patches of the sky are considered. In matrix notation, Eq.~(\ref{eq:start}) can be written in the form
\begin{equation} \label{eq:model}
\Yb = \Ab \Sb, 
\end{equation}
where
\begin{equation}
\Sb= 
\left( \begin{array}{c}
\Sb_1 \\
\Sb_2 \\
\vdots \\
\Sb_{N_c}
\end{array} \right),
\end{equation}
and
\begin{equation}
\Ab = 
\left( \begin{array}{cccc}
a_{11} & a_{12} & \ldots & a_{1N_c} \\
a_{21} & a_{22} & \ldots & a_{2N_c} \\
\vdots & \vdots & \ddots & \vdots \\
a_{N1} & a_{N2} & \ldots & a_{NN_c}
\end{array} \right),
\end{equation}
denotes the so called {\it mixing matrix}.

Here the problem is that only the mixtures $\Yb$ are available, whereas neither $\Ab$ nor $\Sb$ are known. Hence, the issue raises whether 
from $\Yb$ it is possible to obtain the components $\{ \Sb_i \}$. Surprisingly, a positive answer is possible.
At this point, however, it is necessary to stress that the problem presents 
a basic ambiguity. In particular, at best each $\Sb_i$ can be determined unless a multiplicative constant. In fact, if $\Sb_i$ is multiplied by 
a scalar and the corresponding $i$-th column of $\Ab$ is divided by same quantity, then an identical model is obtained. 
For this reason, it is customary to assume that the variance of $\Sb_i$ is equal to one, i.e., $\Sb_i \Sb_i^T / M = 1$.

For simplicity, in the following it is assumed that $N_c = N$, i.e. that the number of observed mixtures is equal to the number of components. 
In this way, $\Ab$ is a square matrix.

\section{A CLASSIC APPROACH: ICA}

One of the most celebrated technique for the blind separation of signals in mixtures is the so called {\it independent component analysis} (ICA).
The basic idea behind ICA is rather simple (and obvious): to obtain the separation of the components it is sufficient to have an estimate of 
$\Ab^{-1}$. In fact, $\Sb = \Ab^{-1} \Yb$. Now, if the CMB component and the Galactic ones are mutually uncorrelated, i.e.
if the corresponding covariance matrix is given by $\Cb_{\Sb} = \Sb \Sb^T / M = \Ib$, then
\begin{equation} \label{eq:covA}
\Yb \Yb^T / M = \Ab \Ab^T.
\end{equation}
This system of equations defines $\Ab$ but an orthogonal matrix. In fact, if $\Ab = \Zb \Vb$, with $\Vb$
orthogonal, then $\Yb \Yb^T / M = \Ab \Ab^T = \Zb \Vb \Vb^T \Zb^T = \Zb \Zb^T $. The problem is that, given the symmetry of $\Yb \Yb^T$,
system~(\ref{eq:covA}) contains only $N (N+1)/2$ independent equations, but the estimates of $N^2$
quantities should be necessary. In ICA, the $N (N-1)/2$ remaining equations are obtained by enforcing the constraint that the components 
$\{ \Sb_i \}$ are not only mutually uncorrelated but also mutually independent. In other words, the separation problem is converted into the form 
\footnote{We recall that the function ``$\arg\min F(x)$''
provides the values of $x$ of for which the function $F(x)$ has the smallest value.}
\begin{equation} \label{eq:ica1}
\widehat \Sb = \underset{\Sb}{\arg\min} \Fb(\Sb), 
\end{equation}
\begin{equation} \label{eq:ica2}
\textrm{subject to} ~~~ \widehat \Sb=\Ab^{-1} \Yb ~~~ \textrm{and} ~~~ \Yb \Yb^T / M = \Ab \Ab^T,
\end{equation}
with $\Fb(\Sb)$ a function that measures the independence between the components $\{ \Sb_i \}$. The definition of a 
reliable measure $\Fb(.)$ is not a trivial task. In literature, various choices are available \citep[see ][ and reference therein]{hyv01}. In practical 
algorithms, the optimization problem is not implemented explicitly in the form~(\ref{eq:ica1})-(\ref{eq:ica2}). Typically, a first estimate
$\widehat \Sb_*$ is obtained through a {\it principal component analysis} (PCA) step followed by a {\it sphering} operation 
(i.e. forcing $\widehat \Sb_* \widehat \Sb_*^T / M = 1$). In this way,
a set of uncorrelated components become available with $\Cb_{\widehat \Sb_*} = \Ib$, as well the corresponding mixing matrix  $\Ab_*$.
Later, these estimated quantities are iteratively refined to maximize $\Fb(\Sb)$ and to get the final $\widehat \Sb$.
Again, in literature, various techniques are available. Among these, one of the most famous and used algorithm is {\it FASTICA} based on a
{\it fixed-point} optimization approach \citep{hyv01,mai02,bac04}. An alternative technique is {\it JADE} that makes use of the 
{\it joint diagonalization algorithm} \citep{car99}.

\section{A SUBSPACE APPROACH} \label{sec:sub}

The main benefit of ICA is the fact that, apart from the independence of the components $\{ \Sb_i \}$, it does not make use of further assumptions.
If from one side, this makes the method easy to use, on the other one it does not permit the exploitation of any information that
be available in advance. In particular, in the case of the CMB observations, it is expected that, at least on small patches of the sky, the mixing matrix 
$\Ab$ can be written in form:
\begin{equation} \label{eq:Am}
\Ab = 
\left( \begin{array}{cccc}
1 & a_{12} & \ldots & a_{1N} \\
1 & a_{22} & \ldots & a_{2N} \\
\vdots & \vdots & \ddots & \vdots \\
1 & a_{N2} & \ldots & a_{NN}
\end{array} \right).
\end{equation}
This means that the component $\Sb_1$, here assumed to correspond to the CMB emission, gives the same contribution in all the observed mixtures. 
Here, the question is how to implement this piece of information. A possible solution can be obtained if model~(\ref{eq:model}) is written in the form:
\begin{equation} \label{eq:basicm}
\Yb =  
\left( \begin{array}{c}
1 \\
1 \\
\vdots \\
1 
\end{array} \right) \Sb_1 +
\left( \begin{array}{c}
a_{12} \\
a_{22} \\
\vdots \\
a_{N2} 
\end{array} \right) \Sb_2 + \cdots +
\left( \begin{array}{c}
a_{1N} \\
a_{2N} \\
\vdots \\
a_{NN} 
\end{array} \right) \Sb_N.
\end{equation}
This equation enlightens the fact that the columns of $\Ab$ span  a {\it signal space} $< \Ab > = {\rm span}\{ \oneb, \ab_2, \ldots, \ab_N \}$
where $\onesb = (1, 1, \ldots, 1)^T$ and $\ab_i$ denotes the i-th column of $\Ab$. In other words, signal $\Yb$ lives in a 
$N$-dimensional space. Moreover, from the same equation it is evident that if $\Yb$ is projected onto a $(N-1)$-dimensional subspace
orthogonal to vector $\oneb$, then the contribution of $\Sb_1$ is removed from $\Yb$ itself. This job can be done by means of the projection matrix
\begin{equation}
\Pb_{\perp} = \Ib - \onesb (\onesb^T \onesb)^{-1} \onesb^T. 
\end{equation}
The application of ICA to
\begin{equation}
\Yb_{\perp} = \Pb_{\perp} \Yb,
\end{equation}
permits to obtain the estimates $\Shb_2, \Shb_3, \cdots, \Shb_N$ of
the $N-1$ components $\Sb_2, \Sb_3, \cdots, \Sb_N$. Hence, since it is $\{ \Shb_i \Shb_j^T / M \}_{i,j=2}^N= \delta_{ij}$ ($\delta_{ij}$ 
denotes the Kronecker function)
and by assumption $\{ \Shb_1 \Shb_i^T \}_{i=2}^N = 0$, the columns of $\Ab$ can be estimated through Eq.~(\ref{eq:basicm}) by means of 
\begin{equation} \label{eq:acoeff}
\ahb_i = \Yb \Shb_i^T / M,~~~~ i=2, 3, \ldots, N.
\end{equation} 
At this point, always using Eq.~(\ref{eq:basicm}), $\Shb_1$ can be derived from
\begin{equation}
\Shb_1 = \onesb^T (\Yb - \Ahb_{-1} \Shb_{-1}) / N, \label{eq:S1} \\
\end{equation}
where
\begin{equation}
\Ahb_{-1} = (\ahb_2, \ldots, \ahb_N),
\end{equation}
and
\begin{equation}
\Shb_{-1}= 
\left( \begin{array}{c}
\Shb_2 \\
\vdots \\
\Shb_{N}
\end{array} \right).
\end{equation}
Since $\Shb_1 \Shb_i^T = 0$, $i\neq 1$, it is not difficult to see that
\begin{equation} \label{eq:fsol}
\Shb_1 = \Sb_1.
\end{equation}
It is worth noticing that, contrary to the other components, the variance of $\Shb_1$ is not one: $\Shb_1 \Shb_1^T / M \neq 1$.
This is a consequence of the fact that $\ab_1 \equiv \onesb$ is fixed.

Here, it is necessary to stress that, if one is interested only in the component $\Sb_1$, then the situation is simpler since it is not
necessary that the other components are independent of even uncorrelated. The point is that
the separation of the components $\{ \Sb_i \}_{i=2}^N$ has no effect on the computation of $\Sb_1$. In other words,
it does no matter whether these components are correctly disentangled or not. In fact, the same $\Shb_1$ as in Eq.~(\ref{eq:S1}) is obtained if,
instead of the independent $\{ \Shb_i \}_{i=2}^N$, in Eq.~(\ref{eq:acoeff}), the orthogonal $\{ \Shb_i^{\star} \}_{i=2}^N$ are used that are 
computed through the application of the PCA to $\Yb_{\perp}$. 
This is because Eq.~(\ref{eq:acoeff}) requires the orthogonality of the components, not their independence. Moreover, 
Eq.~(\ref{eq:model}) can be written in the equivalent form 
\begin{equation} \label{eq:shb}
\Yb = \widetilde \Ab \widetilde \Sb,
\end{equation}
where
\begin{equation}
\widetilde \Ab = \Ab \Hb,
\end{equation}
\begin{equation}
\widetilde \Sb = \Hb^{-1} \Sb,
\end{equation}
\begin{equation}
\Hb = 
\left( \begin{array}{ccccc}
1 & \vline & 0 & \ldots & 0 \\
\hline
0 & \vline & & &  \\
\vdots & \vline & & \Phib & \\
0 & \vline & & &
\end{array} \right),
\end{equation}
and  $\Phib$ is any arbitrary $(N-1) \times (N-1)$ non-singular matrix. Now, since $\widetilde \Sb_1 = \Sb_1$ and the first colum of $\widetilde \Ab$
is still $\onesb$, the meaning of this equation is that there is an infinite number of sets $\{ \widetilde \Sb_i \}_{i=2}^N$ that, 
when projected onto the $N-1$ subspace orthogonal to $\onesb$, produce the same $\Yb_{\perp}$. 
Hence, for the separation of $\Sb_1$ it is not necessary the use of the true $\{ \Sb_i \}_{i=2}^N$ but only one of such sets.

Of course, these are theoretical results. In practical situations, it is quite improbable that for finite signals
the condition $\{ \Sb_1 \Sb_i^T \}_i= 0$, $i \neq 1$ be strictly satisfied. In fact, because of the statistical fluctuations, in general 
the components $\{ \Sb_i \}_{i=1}^N$ present a certain degree of mutual correlation even in the case they are the realization of independent, stationary, 
stochastic random processes. Therefore, enforcing the condition $\Shb_1 \Shb_i = 0$, $i\neq 1$,
makes inaccurate the estimate of the coefficients $\{ \ahb_i \}$ as provided by Eq.~(\ref{eq:acoeff}).

\section{SOME NUMERICAL EXPERIMENTS} \label{sec:experiments}

Because of the arguments presented above, it is to be expected that, with respect to the classic ICA, the use of 
a subspace method provides more accurate and stable results. Given the non-linear nature of the algorithms, the verification
of this expectation has to be made through numerical experiments.

Here, non-astronomical subjects have been deliberately chosen. In this way,
a direct visualization of the separation is possible and hence an easier and safer assessment of its goodness. Moreover, 
the use of {\it deterministic} subjects make easier the modeling of various experimental conditions (e.g., the sample correlation
between different images can be obtained using images with almost-constant luminosity areas in correspondence to the 
same coordinates).

Figures~\ref{fig:good1}-\ref{fig:good2} show the results provided by ICA and its version based on the subspace approach (modified ICA)
when three mixtures are available each containing the contribution of an equal number of (almost) uncorrelated components
with a mixing matrix 
\begin{equation} \label{eq:A}
\widehat \Ab = 
\left( \begin{array}{ccc}
1.0 &  0.2 & 0.4 \\
1.0 &  0.5 & 0.3 \\
1.0 &  0.3 & 0.6
\end{array} \right).
\end{equation}
No noise has been added. The examination of these figures seems to indicate that both techniques are able
to produce an excellent separation. Actually, the results provided by the classic ICA are unstable and often unsatisfactory. The point is that
this method is non-linear and therefore the algorithms have to be initialized. As a consequence, different results
can be obtained according to the chosen initialization. This is quantified in Fig.~\ref{fig:good3}, where the normalized norm of
the residuals, $|| \Shb_i - \Sb_i || / || \Sb_i||$, corresponding to $200$ different initializations are shown.
The residuals are computed from the difference between the real solution and the estimated one. From this figure, it is clear that the classical 
ICA provides more more than one solution, while the modified ICA method produces a stable $\Shb_1$. This is a consequence of the fact that in the subspace 
approach such component is computed via linear operations only.

The question of the stability of the solution may be considered of secondary
importance. Actually, this is not true. The ICA method does not permit to check if a specific
solution is satisfactory or not. Any separation provides components that, when summed up, will perfectly reproduce the original
mixtures. In a real experimental situation, this means the unavailability of a reliable selection criterion. For example, a simple selection
criterion could be based on the frequency with which a solution is obtained for different initializations of the algorithm. However, 
there is no guarantee that the most frequent solution represents the best one. This is the case for the experiment in
Figs.~\ref{fig:bad1}-\ref{fig:bad3} where, contrary to the previous one, the components present a certain degree of correlation with a 
(normalized) cross-product matrix
\begin{equation} \label{eq:cross}
\widehat \Cb_{\Sb} = 
\left( \begin{array}{ccc}
1.00 &  0.20 & 0.06 \\
0.20 &  1.00 & 0.06 \\
0.06 &  0.06 & 1.00
\end{array} \right).
\end{equation}
As expected, the separation is by far less satisfactory than in the previous experiment. However, also in this case the results 
concerning the subspace method appear more stable. Here, the point of interest is that from the examination of the top-left panel 
in Fig.~\ref{fig:bad3} 
it is possible to see that the classic ICA provides the estimate $\Shb_1$ closest to the true solution. However, apart from the fact that 
the $\Shb_2$ corresponding to such estimate is systematically the worst among those obtained, the frequency with which the ``{\it best}''
$\Shb_1$ is found is very small. In a practical situation, such a solution should have been discarded. 

It has not to surprise that, when the conditions of applicability are violated, the classic ICA method is able to ``see'' a good estimate of $\Sb_1$ 
that is ``unreachable'' with the subspace approach. This is because, as stated above, the classic ICA works with a larger number of unknowns and therefore 
is more flexible and can span a wider ``solution space''. However, there is no guarantee that such a flexibility is effectively 
fruitful. The situation is similar (although not identical) to a polynomial fit: the use of high degree functions permits a greater flexibility but 
at the cost of a remarkable instability of the results that often make quite hard, if not impossible, the choice of a good solution.    

\section{RELATIONSHIP WITH THE ILC METHOD} \label{sec:ilc}

In CMB literature, another method has been often used for the separation of Cosmic signals from the Galactic foregrounds.
This is the so called {\it internal linear combination} method (ILC) \citep{ben03, eri04, hin07}. The aim of this method is not the separation of all the
signals that contribute to the observed mixtures, but only the extraction of the specific component $\Sb_1$.
With ILC a solution is searched in the form
\begin{equation} \label{eq:ilc}
\Shb_1 = \wb^T \Yb,
\end{equation} 
with the column vector $\wb$ providing a set of appropriate weights. Since, the basic assumption is that $\Sb_1$ is the same in all the mixtures,
i.e.,
\begin{equation} \label{eq:Y}
\Yb = \onesb \Sb_1 + \Nmthb;
\end{equation}
with $\Nmthb$ a zero-mean noise that provides the contribution of all the components other than that of interest, it is imposed that
\begin{equation} \label{eq:w}
\onesb^T \wb = 1. 
\end{equation}
In this way,
\begin{equation}
\Shb_1 = \Sb_1 + \wb^T \Nmthb,
\end{equation} 
i.e. the weights do not alter the $\Sb_1$ component.
For the same assumption, among all the possible solutions provided by Eq.~(\ref{eq:ilc}) with the condition~(\ref{eq:w}), that of interest
has the property that $\sigma^2 = \Shb_1 \Shb^T_1$ is a minimum. In fact, assuming the noise $\Nmthb$ uncorrelated with $\Sb$, it is
\begin{equation} \label{eq:sigma}
\sigma^2 = \frac{1}{M} [\Sb_1 \Sb_1^T + \wb^T \Nmthb \Nmthb^T \wb].
\end{equation}
Hence, the minimization of $\sigma^2$ with respect to $\wb$ implies the strongest filtering of the component $\Nmthb$.
It can be shown that the weights which minimize this quantity are given by \citep{eri04}
\begin{equation} \label{eq:wr}
\wb = \frac{\Cb_{\Yb}^{-1} \oneb}{\oneb^T \Cb_{\Yb}^{-1} \oneb},
\end{equation}
where 
$\Cb_{\Yb} = \Yb \Yb^T / M$. Hence, the ILC estimator takes the form
\begin{equation} 
\Shb_1 = \frac{\oneb^T \Cb_{\Yb}^{-1} \Yb}{\oneb^T \Cb_{\Yb}^{-1} \oneb}. \label{eq:basic}
\end{equation}
Although this estimator appears different from that given by Eq.~(\ref{eq:S1}), actually they provide identical results.
In fact, under model~(\ref{eq:basicm}), it is
\begin{equation} \label{eq:cov}
\Cb_{\Yb} = \Ab \Cb_{\Sb} \Ab^T.
\end{equation}
If this equation is inserted in Eq.~(\ref{eq:basic}), one obtains
\begin{equation} \label{eq:solution}
\Shb_1 = \alpha \oneb^T \Ab^{-T} \Cb_{\Sb}^{-1} \Sb,
\end{equation}
with the scalar $\alpha$ given by
\begin{equation}
\alpha = [ \oneb^T \Ab^{-T} \Cb_{\Sb}^{-1} \Ab^{-1} \oneb]^{-1},
\end{equation}
and $\Ab^{-T} \equiv (\Ab^{-1})^T$.
Now, since it is trivially verified that 
\begin{equation}
\oneb^T = \eb_1^T \Ab^T,
\end{equation}
where
\begin{equation} \label{eq:trick2}
\eb_1 \equiv (1, 0, \ldots, 0)^T,
\end{equation}
it is
\begin{equation} \label{eq:trick1}
\oneb^T \Ab^{-T} = \eb_1^T.
\end{equation}
Hence, $\alpha = (\Cb_{\Sb}^{-1})_{1 1} = ([\Sb_1 \Sb_1^T])^{-1} = \sigma_{11}^{-1}$.
As a consequence, if $\Sb_1$ is uncorrelated with $\{ \Sb_i \}_{i=2}^N$ 
i.e. if
\begin{equation}
\Cb_{\Sb} = 
\left( \begin{array}{cccccc} \label{eq:covar1}
\sigma_{11} & \vline & 0 & 0 & \ldots & 0 \\
\hline
0 & \vline & \sigma_{22} & \sigma_{23} & \ldots & \sigma_{2 N} \\
\vdots & \vline & \vdots & \vdots & \ddots & \vdots \\
0 & \vline & \sigma_{N 2} & \sigma_{N 3} & \ldots & \sigma_{N N}
\end{array} \right),
\end{equation}
from Eq.~(\ref{eq:solution}) and the fact that $\Cb_{\Sb}^{-1}$ has the form
\begin{equation}
\Cb^{-1}_{\Sb} = 
\left( \begin{array}{cccccc}
\sigma^{-1}_{11} & \vline & 0 & 0 & \ldots & 0 \\
\hline
0 & \vline & & & & \\
\vdots & \vline & & & \Omegab^{-1} & \\
0 & \vline & & & &
\end{array} \right),
\end{equation}
with $\Omegab$ the bottom-right block of the matrix in the rhs of Eq.~(\ref{eq:covar1}), one obtains that 
\begin{equation}
\Shb_1 = \eb_1^T \Sb = \Sb_1,
\end{equation}
i.e., the same result as Eq.~(\ref{eq:fsol}). More in general, the 
estimators~(\ref{eq:S1}) and (\ref{eq:solution}) provide identical results also
when the components $\Sb_1$ is not uncorrelated with the other ones and/or instrumental noise is added to the observed
mixtures. The reason is that, as stated earlier, ILC provides an estimate $\Shb_1$ with the property that the quantity $\sigma^2$ in 
Eq.~(\ref{eq:sigma}) is 
a minimum. Although not evident from the treatment in Sec.~\ref{sec:sub}, the same holds for the estimator~(\ref{eq:S1}). 
In fact, it is not difficult to realize that, after the determination of the components $\{ \Shb_i \}_{i=2}^N$, the coefficients $\{ \ahb_i \}_{i=2}^N$,
as given by Eq.~(\ref{eq:acoeff}), are the solution of
\begin{equation}
\frac{d \sigma^2}{d \Ahb_{-1}} = \frac{d (\Shb_1 \Shb_1^T)}{d \Ahb_{-1}} = 0
\end{equation}
with
\begin{equation}
\Shb_1 \Shb_1^T = \onesb^T (\Yb - \Ahb_{-1} \Shb_{-1}) (\Yb - \Ahb_{-1} \Shb_{-1})^T \onesb
\end{equation}
that is derived from the sample version of Eq.~(\ref{eq:basicm}).

\section{FINAL REMARKS}

In the previous section it has been assumed that the number $N$ of the observed mixtures (or frequency channels, images at different observing frequency) 
equals the number $N_c$ of the components. 
In practical application this coincidence is improbable. Of course, in order the subspace approach can work satisfactorily, it is necessary
to know the correct dimension of the signal-space. Therefore, the question raises on what happens when $N_c \neq N$. If from one side, the case
$N < N_c$ does not offer many possibilities to obtain meaningful results, on the other one the case $N > N_c$ can be successfully addressed.
In fact, $N_c$ can be determined by the number of non-zero eigenvalues of matrix $\Cb_{\Yb}$ and the corresponding eigenvectors
can be used to construct a basis of the signal-space with the correct dimensionality. After that, it is sufficient to project
$\Yb$ onto this space, obtaining a ``new'' set of $N_c^* < N$ mixtures $\Yb_{R}$, and then to work with this. The rest of the procedure in 
Sect.~\ref{sec:sub} and Sec.~\ref{sec:ilc} remains the same.
More in particular, if the $N \times N$ matrix $\Cb_{\Yb}$ is decomposed in the form
\begin{equation}
\Cb_{\Yb} = \Vb \Lambdab \Vb^T
\end{equation} 
where $\Lambdab$ is a diagonal matrix containing the eigenvalues $\lambda_1 \ge \lambda_2 \ge \ldots \ge \lambda_{N_c} > 
\lambda_{N_c+1} = \ldots = \lambda_N = 0$, whereas $\Vb$ is an orthogonal matrix whose 
columns contain the corresponding eigenvectors, then
\begin{equation}
\Yb_{R} = \Lambdab^{-1/2}_{\varnothing} \Vb^T \Yb, 
\end{equation}
where $\Lambdab^{-1/2}_{\varnothing}$ is a diagonal matrix whose non-zero values are given by the inverse of the non-zero entries of $\Lambdab^{1/2}$.

More complex is the situation when $\Yb$ is contaminated by measurements errors $\Nb$. If $\Nb$ is zero-mean and additive, model~(\ref{eq:model}) 
converts into
\begin{equation} \label{eq:modelN}
\Yb = \Ab \Sb + \Nb.
\end{equation}
Here the problem is that the effective number of components becomes $N_c + N$ and therefore the separation problem is always underdetermined. 
Therefore, although ${\rm E}[\Nb] = \zerosb$, it happens that $E[\Shb_1] \neq \Sb_1$. In other words, a bias is present \citep{vio08}. 
This point is more evident
if the ILC estimator~({\ref{eq:basic}) is considered. There, the bias is due to the fact that matrix $\Cb_{\yb}$ is no longer given by
Eq.~(\ref{eq:cov}) but by 
\begin{equation} \label{eq:covN}
\Cb_{\Yb} = \Ab \Cb_{\Sb} \Ab^T + \Cb_{\Nb}.
\end{equation}
This equation suggests that a simple way to remove the bias is to use $\Cb_{\Yb} - \Cb_{\Nb}$ instead of $\Cb_{\Yb}$.
Something similar holds also for $\Shb_1$ as provided by Eq.~(\ref{eq:S1}). In fact, after some algebra it is possible to show
that an equivalent form is
\begin{equation}
\Shb_1 = \onesb^T [\Ib - \Cb_{\Yb} \Pb_{\perp}^T (\Pb_{\perp} \Cb_{\Yb} \Pb_{\perp}^T)^{\dagger} \Pb_{\perp} ] \Yb / N,
\end{equation}
where symbol ``$~{}^{\dagger}~$'' denotes {\it pseudo-inverse}.
Hence, the same arguments apply as above. Concerning the influence of the noise of the components $\{ \Sb_i \}$, $i \neq 1$, the situation is much
more difficult since similar to that encountered in the classic ICA approach \citep[for details, see][]{hyv01}.

\section{SUMMARY}

In this paper we have considered the problem of a modification of the ICA separation technique that permits to exploit the ``{\it a priori}''
information the, contrary to the Galactic components, the contribution of CMB to the microwave maps is independent of the observing frequency.
A subspace approach has been proposed that is more stable and provide more accurate results than the classic ICA technique. A relationship between 
this approach and the {\it Internal Linear Composition} method has been also shown.

\clearpage
\begin{figure*}
        \resizebox{\hsize}{!}{\includegraphics{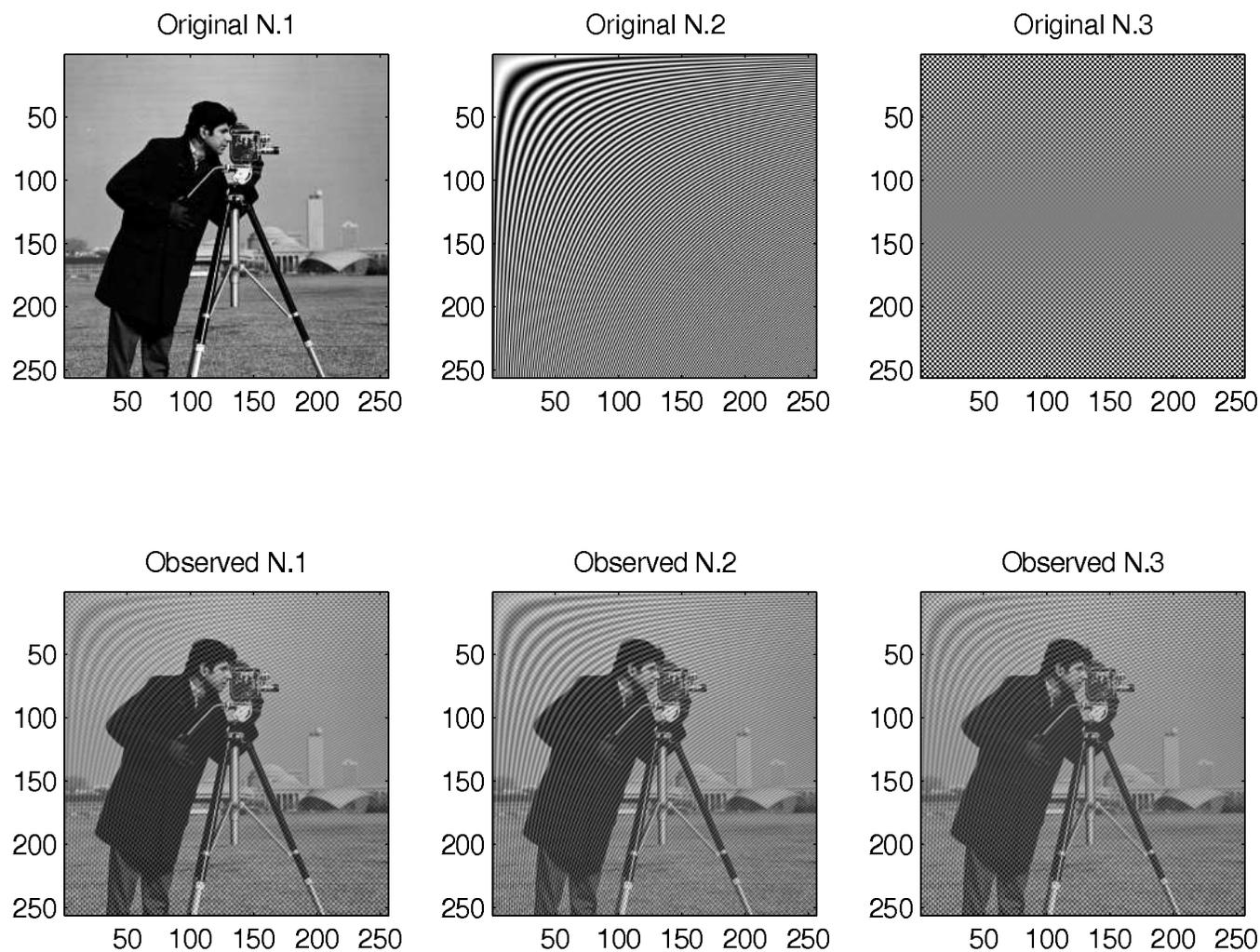}}
        \caption{{\bf Top panels} -- Original images $\Sb$ used in the experiment dealing with component separation as described
        in Sect.~\ref{sec:experiments}; {\bf Bottom panels} -- Observed mixtures 
        $\Yb = \Ab \Sb$, with $\Ab$ given by Eq.~(\ref{eq:A}). In this experiment the components are almost uncorrelated.}
        \label{fig:good1}
\end{figure*}
\clearpage
\begin{figure*}
        \resizebox{\hsize}{!}{\includegraphics{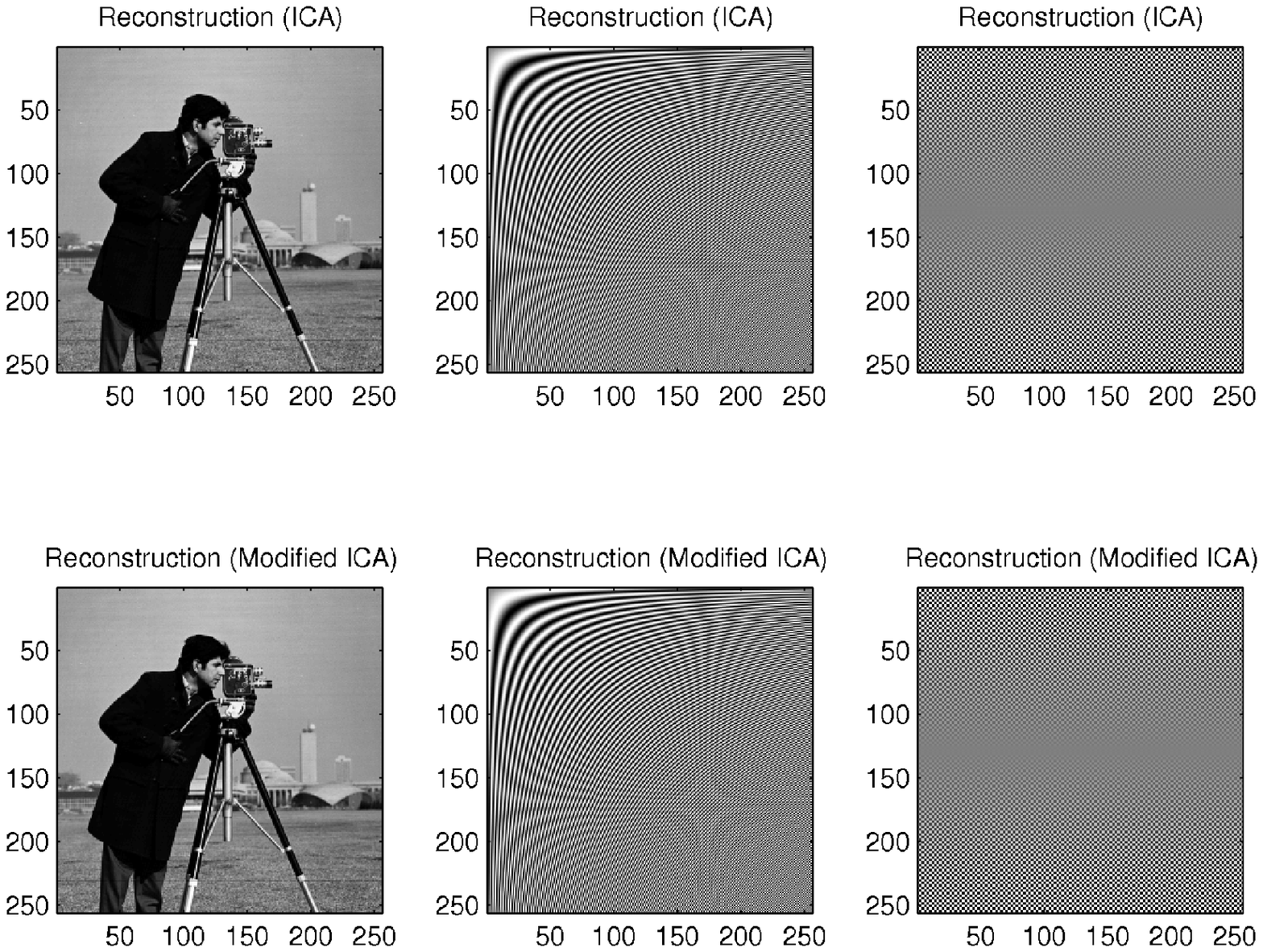}}
        \caption{{\bf Top panels} -- Typical separation obtained with the classic ICA algorithm when applied to the observed mixtures shown
         in Fig.~\ref{fig:good1};
        {\bf Bottom panels} -- Typical separation obtained with the subspace based ICA when applied to the observed mixtures shown in
        Fig.~\ref{fig:good1}.}
        \label{fig:good2}
\end{figure*}
\clearpage
\begin{figure*}
        \resizebox{\hsize}{!}{\includegraphics{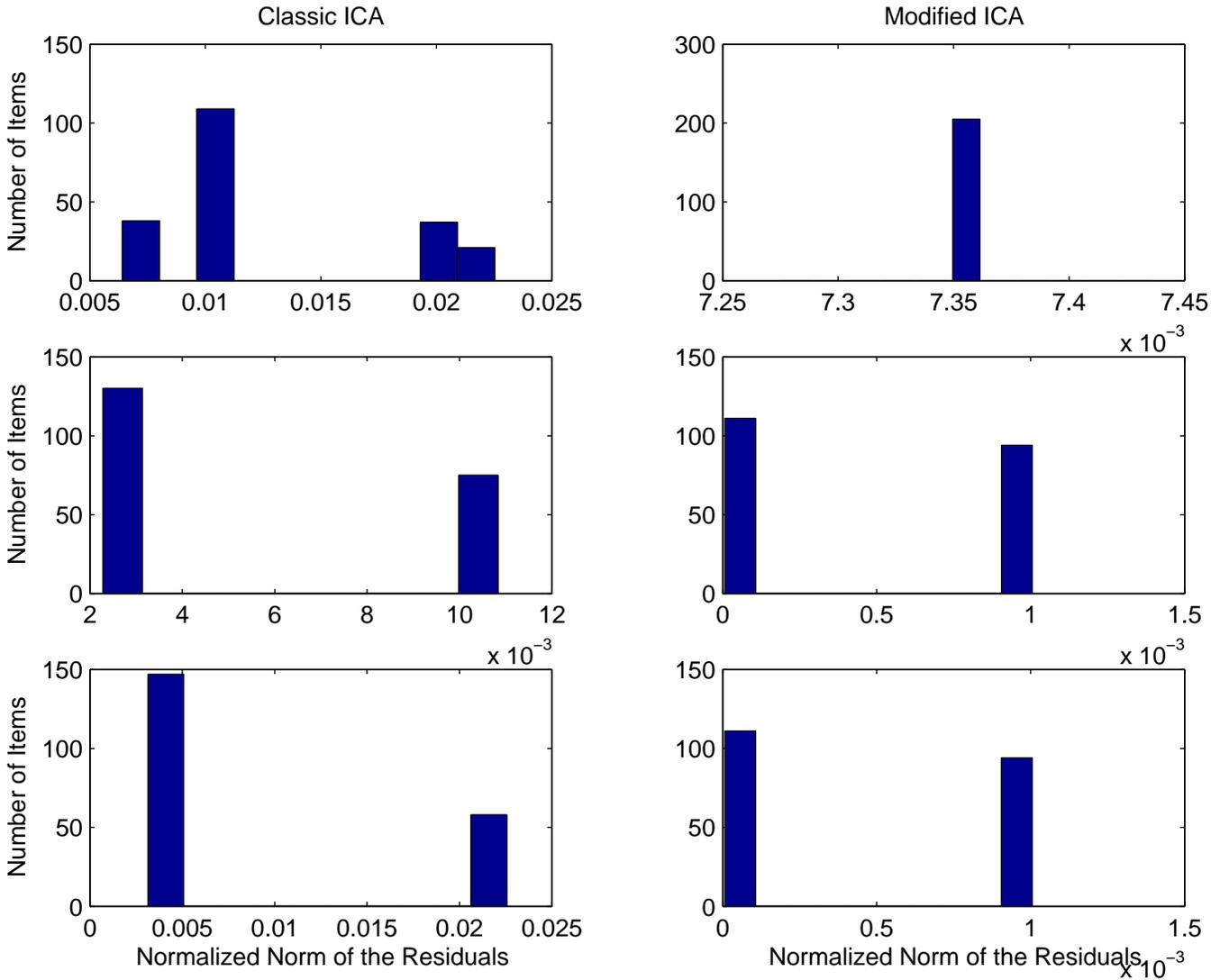}}
        \caption{Histogram of the normalised residuals $|| \Shb_i - \Sb_i || / || \Sb_i||$. The residuals are defined as the difference
        between the real and the estimated solution. The corresponding solutions of the component separations are obtained with $200$ different 
        inizializations of the classic ICA and subspace based ICA methods when applied to the observed mixtures in Fig.~\ref{fig:good1}: 
        top panels correspond to component $i=1$, central panels to $i=2$, bottom panels to $i=3$. As shown in the top panel the solution
        in the modified ICA approach for component $i=1$ is stable.}
        \label{fig:good3}
\end{figure*}

\clearpage
\begin{figure*}
        \resizebox{\hsize}{!}{\includegraphics{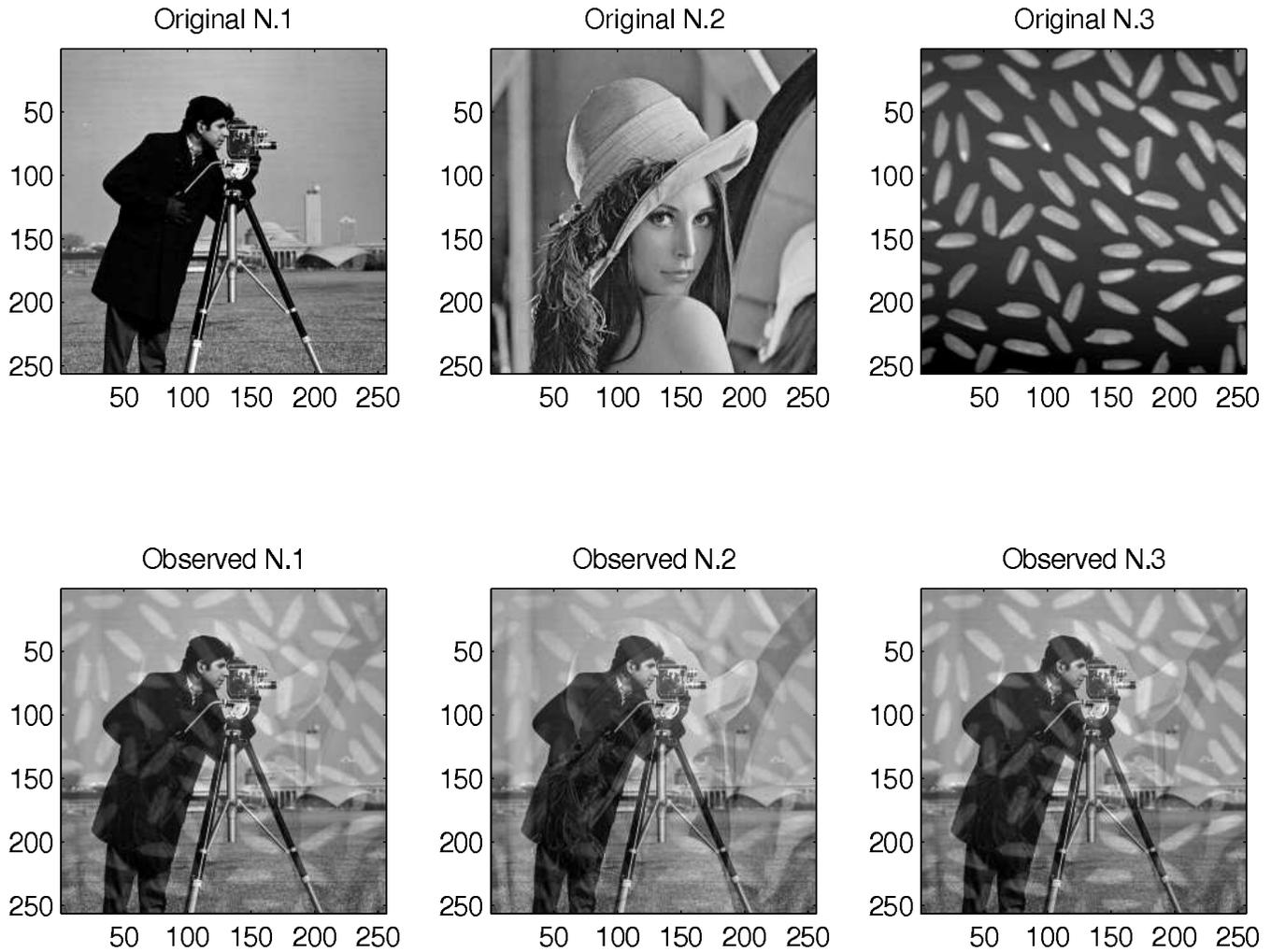}}
        \caption{{\bf Top panels} --  Original images $\Sb$ used in the experiment dealiong with component separation as
        described in Sect.~\ref{sec:experiments}; {\bf Bottom panels} -- Observed mixtures 
        $\Yb = \Ab \Sb$. In this experiment the components are partially correlated with the correlation matrix given by Eq.~(\ref{eq:cross}). }
        \label{fig:bad1}
\end{figure*}
\clearpage
\begin{figure*}
        \resizebox{\hsize}{!}{\includegraphics{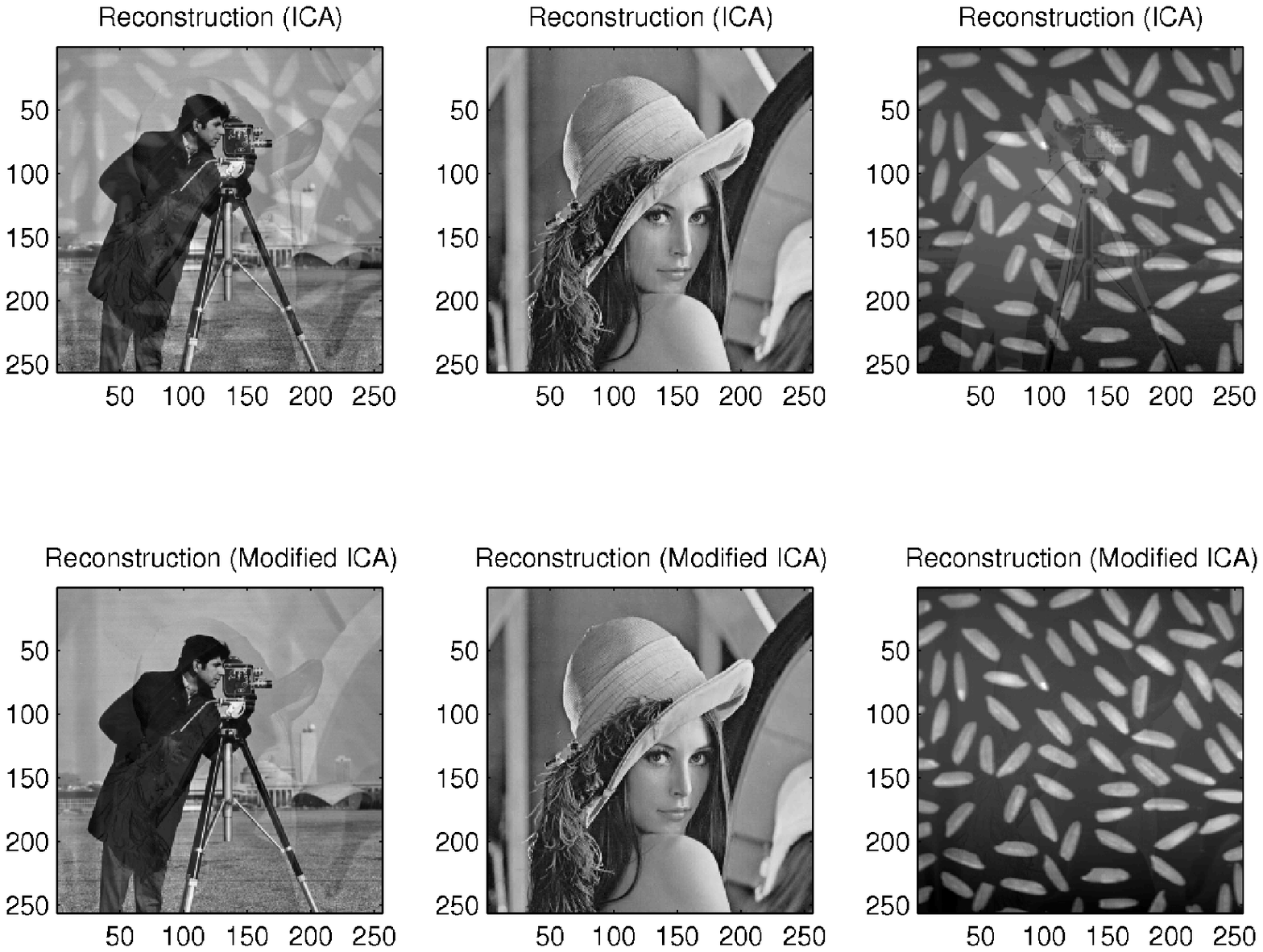}}
        \caption{{\bf Top panels} -- Typical separation obtained with the classic ICA algorithm when applied to the observed mixtures in 
        Fig.~\ref{fig:bad1};
        {\bf Bottom panels} -- Typical separation obtained with the subspace based ICA when applied to the observed mixtures in Fig.~\ref{fig:bad1}.}
        \label{fig:bad2}
\end{figure*}
\clearpage
\begin{figure*}
        \resizebox{\hsize}{!}{\includegraphics{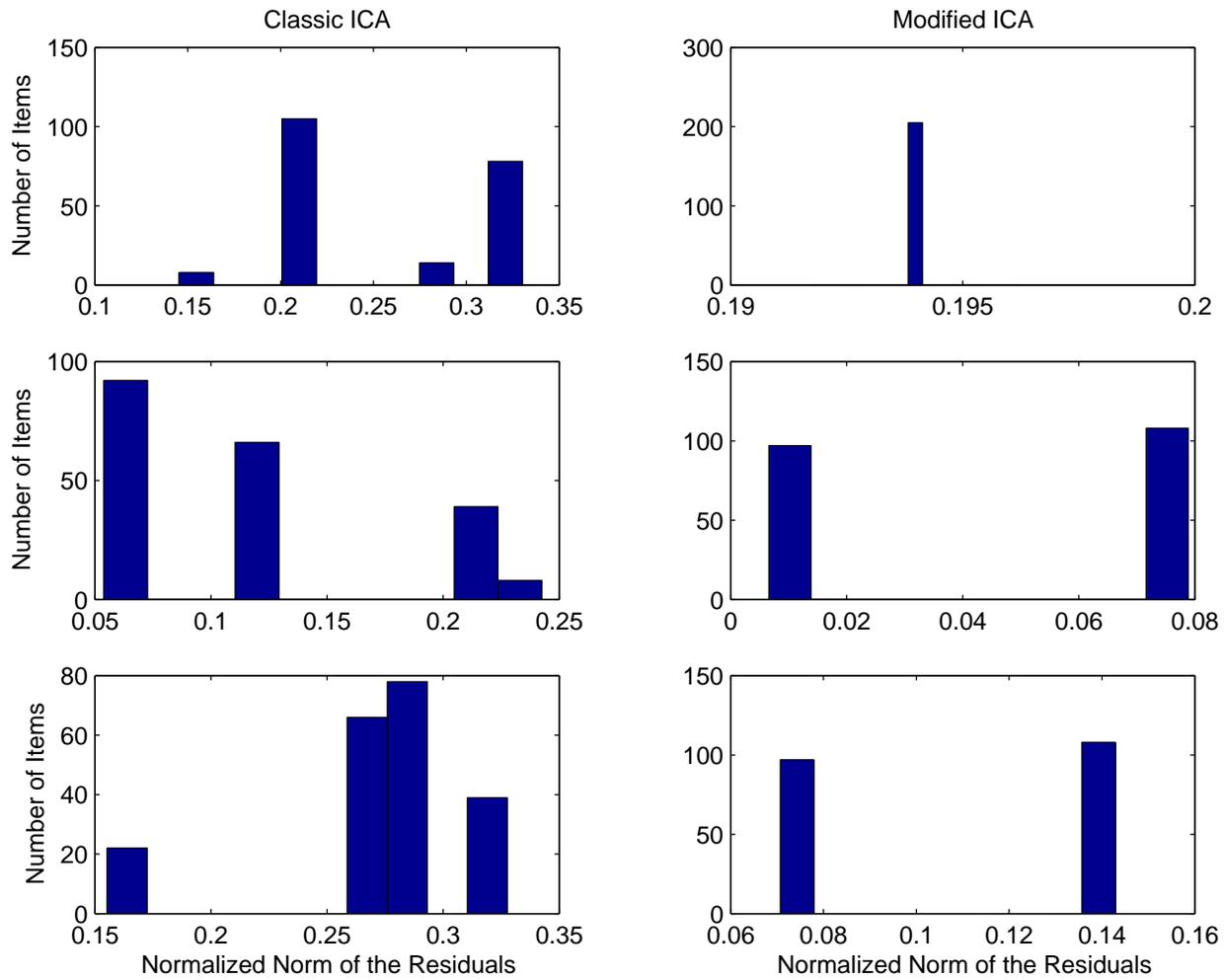}}
        \caption{Same as Figure \ref{fig:good2} for the simulations shown in Figure \ref{fig:bad1} and \ref{fig:bad2}.}
        \label{fig:bad3}
\end{figure*}


\begin{thebibliography}{}
\bibitem[Baccigalupi et al. (2004)]{bac04} Baccigalupi, C., et al. 2004, \mnras, 354, 55
\bibitem[Bennett et al. (2003)]{ben03} Bennett, C.L., et al. 2003, \apjs, 148, 97  
\bibitem[Cardoso (1999)]{car99} Cardoso, J.F. 1999, Neural Computation, 11, 157
\bibitem[Eriksen et al. (2004)]{eri04} Eriksen, H.K., Banday, A.J., G\'orski, K.M., \& Lilje, P.B. 2004, \apj, 612, 633
\bibitem[Hinshaw et al. (2007)]{hin07} Hinshaw, G., et al. 2007, \apjs, 170, 288
\bibitem[Hyv\"arinnen et al. (2001)]{hyv01} Hyv\"arinnen, A., Karhunen, J., \& Oja, E. 2001, Independent Component Analysis (New York: John Wiley \& Sons)
\bibitem[Maino (2002) et al. 2002)]{mai02} Maino, D., et al. 2002, \mnras, 334, 53
\bibitem[Stivoli et al. (2006)]{sti06} Stivoli, F., Baccigalupi, C., Maino, D., \& Stompor, R. 2006, \mnras 372, 615
\bibitem[Vio \& Andreani (2008)]{vio08} Vio, R., \& Andreani, P. 2008, A\&A, {\it submitted}
\end{thebibliography}
\end{document}